\documentclass[showpacs,showkeys,twocolumn,nofootinbib,longbibliography]{revtex4-1}

\usepackage{amsmath}
\usepackage{amssymb}
\usepackage[dvips]{graphicx}
\usepackage{epsf}
\usepackage{slashed}
\usepackage{enumitem}
\usepackage[czech,english]{babel}
\usepackage[cp1250]{inputenc}
\usepackage{hyperref}
\usepackage{accents} 

\usepackage[only,llbracket,rrbracket,llparenthesis,rrparenthesis]{stmaryrd} 
\usepackage{accsupp} 

\newcommand{\Tr}{\mathop{\rm Tr}\nolimits}
\newcommand{\I}{\ensuremath{\mathrm{i}}}
\renewcommand{\d}{\ensuremath{\mathrm{d}}}
\newcommand{\hc}{\ensuremath{\mathrm{h.c.}}}
\newcommand{\qm}[1]{``#1''} 

\newcommand{\T}{\ensuremath{\mathrm{T}}}
\newcommand{\C}{\ensuremath{\mathrm{c}}}
\newcommand{\MeV}{\ensuremath{\mathrm{\,MeV}}}

\newcommand{\group}[1]{\mathbb{#1}}

\usepackage{bbm} 

\begin{document}

\title{Contribution of right-handed neutrinos and standard fermions to $W^\pm$ and $Z$ masses}

\author{Petr Bene\v{s}}
\affiliation{Institute of Experimental and Applied Physics, Czech Technical University in Prague, Horsk\'{a} 3a/22, 128\,00 Prague 2, Czech Republic}
\email{p.benes@utef.cvut.cz}

\begin{abstract}
We present expressions of the Pagels--Stokar type for the masses of the $W^\pm$ and $Z$ bosons in terms of the quark and lepton self-energies. By introducing a genuine new term in the gauge boson--fermion--anti-fermion vertex we manage to accomplish three main achievements: First, we show that the similar results existing in literature lead, in general, to a non-symmetric gauge boson mass matrix and we fix this flaw. Second, we consider the case of any number of fermion generations with general mixing. Third, we include in our analysis also an arbitrary number of right-handed neutrinos, together with the left-handed and right-handed neutrino Majorana masses (self-energies). On top of that, we give also a correction to the original Pagels--Stokar formula for the pion decay constant in QCD.
\end{abstract}




\maketitle

\section{Introduction}

The mechanism of spontaneous symmetry breaking is usually assumed to be triggered either by condensation of elementary scalars or by fermion condensates (or more generally, by their dynamically generated masses or self-energies). The latter is the mechanism of chiral symmetry breaking in QCD and is assumed to happen also, e.g., in the Technicolor theories.


Similar mechanism can be in principle responsible also for the electroweak symmetry breaking in the Standard Model (SM). Indeed, if, by means of some dynamics beyond SM, the quarks and leptons obtain masses, they will inevitably break the electroweak symmetry. (The SM Higgs sector is in this picture assumed to be low energy description of the new dynamics.) This idea dates back to the old top-condensations models \cite{Hosek:1985jr,Miransky:1988xi,Bardeen:1989ds} and has been alive ever since.

If the electroweak symmetry is broken, the $W^\pm$ and $Z$ bosons must obtain masses, proportional to the particular order parameters, i.e., in our case the dynamically generated quark and lepton self-energies. In this letter we present explicit formulae for these masses in terms of the self-energies. We take into account not only the SM fermions, but, motivated by recent studies \cite{Hosek:2009ys,Smetana:2011tj}, also the right-handed neutrinos.


\section{Review of traditional approach}
\label{sec_review}

Consider a theory with Abelian axial $\group{U}(1)_{\mathrm{A}}$ gauge symmetry, containing a single\footnote{Ignoring the axial anomaly does not affect the reasoning in this section.} fermion flavor $\psi$, charged under it. The corresponding symmetry generator is defined with the gauge coupling constant $g$ deliberately included: $T_{\mathrm{A}} = g \gamma_5$.

Assume further that the fermion obtains somehow (by means of the $\group{U}(1)_{\mathrm{A}}$ dynamic itself or by some other dynamics) the proper self-energy $-\I \boldsymbol{\Sigma}_p = \langle \psi \bar\psi \rangle_{\mathrm{1PI}}$. Let us for the simplicity make two technical assumptions: First, that $\boldsymbol{\Sigma}_p$ is a function only of $p^2$ (and not of $\slashed{p}$) and second, that it is Hermitian in the sense $\boldsymbol{\Sigma}_p^{\phantom{\dag}} = \gamma_0 \boldsymbol{\Sigma}_p^\dag \gamma_0$. These assumptions constrain $\boldsymbol{\Sigma}_p$ to have the form
\begin{eqnarray}
\label{Sigma_form}
\boldsymbol{\Sigma}_p^{\phantom{*}} &=& \Sigma_p^* P_L + \Sigma_p^{\phantom{*}} P_R \,,
\end{eqnarray}
where $\Sigma_p$ is a complex function of $p^2$ and $P_{R,L} = (1\pm\gamma_5)/2$.

Such a non-vanishing $\boldsymbol{\Sigma}_p$ breaks spontaneously the $\group{U}(1)_{\mathrm{A}}$ gauge symmetry and the corresponding gauge boson thus must acquire a mass. This mass can be obtained from the polarization tensor, which is necessarily of the transversal form $\Pi^{\mu\nu}(q) = (g^{\mu\nu}q^2 - q^\mu q^\nu)\Pi(q^2)$. If the form factor $\Pi(q^2)$ develops a pole of the type $1/q^2$, its residue is (in the lowest approximation) just the gauge boson mass squared.

\begin{figure}[t]
\begin{center}
\includegraphics[width=.5\textwidth]{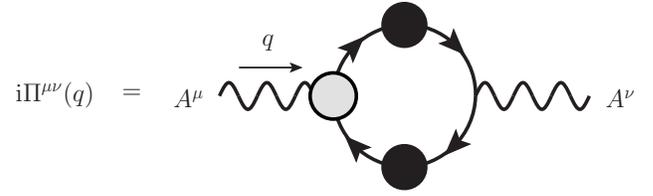}
\caption[Fermion one-loop contribution to $\Pi^{\mu\nu}(q)$.]{The polarization tensor $\Pi^{\mu\nu}(q)$. The full blobs stand for the full propagators, while the gray blob is the proper (1PI) vertex function.}
\label{fig_Pi}
\end{center}
\end{figure}

As argued in \cite{Jackiw:1973tr}, the polarization tensor with the desired pole can be calculated from one-loop fermion contribution, with the propagators given by $G_p = (\slashed{p}-\boldsymbol{\Sigma}_p)^{-1}$, with one insertion of a the bare vertex $\gamma^\mu T_{\mathrm{A}}$ and with the other insertion of the dressed vertex $\Gamma^\mu(p^\prime,p)$ (see Fig.~\ref{fig_Pi}). The point is that $\Gamma^\mu(p^\prime,p)$ must satisfy the Ward--Takahashi (WT) identity 
\begin{subequations}
\label{WT}
\begin{eqnarray}
q_\mu \Gamma^\mu(p^\prime,p) &=& G_{p^\prime}^{-1} T_{\mathrm{A}} - \gamma_0 T_{\mathrm{A}} \gamma_0 G_{p}^{-1}
\\ &=&
g \slashed{q} \gamma_5 - g (\boldsymbol{\Sigma}_{p^\prime}+\boldsymbol{\Sigma}_{p}) \gamma_5 \,,
\end{eqnarray}
\end{subequations}
where $q = p^\prime - p$. The WT identity has two consequences. First, it guarantees transversality of the polarization tensor. Second, it implies that, as long as $\boldsymbol{\Sigma}_p \neq 0$, the vertex $\Gamma^\mu(p^\prime,p)$ must have a pole of the type $q^\mu/q^2$ due to intermediate \qm{would-be} Nambu--Goldstone (NG) boson, corresponding to spontaneous breakdown of the $\group{U}(1)_{\mathrm{A}}$ symmetry. This pole in the vertex gives in turn rise to the pole in $\Pi(q^2)$ and thus to the gauge boson mass.

The last ingredient we need is hence the dressed vertex $\Gamma^\mu(p^\prime,p)$. General principles (especially the WT identity) constrain it to have the form of the bare vertex plus the NG pole part plus a transversal part:
\begin{eqnarray}
\label{vertex}
\Gamma^\mu(p^\prime,p) &=& 
g \gamma^\mu \gamma_5 - g \frac{q^\mu}{q^2} (\boldsymbol{\Sigma}_{p^\prime}+\boldsymbol{\Sigma}_{p}) \gamma_5
+ \tilde\Gamma^\mu(p^\prime,p) \,,
\hspace{2em}
\end{eqnarray}
where $q_\mu\tilde\Gamma^\mu(p^\prime,p)=0$.

In \cite{Pagels:1979hd} it is argued that $\tilde\Gamma^\mu(p^\prime,p)$ can be actually neglected.
The argument is that since $\tilde\Gamma^\mu(p^\prime,p)$ is transversal, it does not contain the NG pole ($\sim q^\mu/q^2$). Therefore it can be safely ignored, because it does not contribute to the gauge boson mass (which is approximated by the residue of the pole of $\Pi(q^2)$).

Now, having the dressed vertex \eqref{vertex} and assuming $\tilde\Gamma^\mu(p^\prime,p) = 0$, the gauge boson mass $m$ can be straightforwardly calculated using the approach described above. The result can be written as $m = g F_\pi$, where the NG boson (\qm{pion}) decay constant $F_\pi$ is
\begin{eqnarray}
\label{pagels_stokar}
F_\pi^2 &=& -2 \I \int\!\frac{\d^4 p}{(2\pi)^4} 
\frac{|\Sigma_p|^2 - \frac{1}{4}p^2|\Sigma_p|^{2\prime}}{(p^2-|\Sigma_p|^2)^2} \,,
\end{eqnarray}
where the prime denotes derivative with respect to $p^2$. This expression for $F_\pi$ is, up to the missing color and flavor number factors, known as the Pagels--Stokar (PS) formula \cite{Pagels:1979hd}.

\section{Novel pole term}
\label{sec_modification}

It is certainly true that transversality of  $\tilde\Gamma^\mu(p^\prime,p)$ does not imply a massless pole. On the other hand, it does not forbid such a pole either. Indeed, we argue that $\tilde\Gamma^\mu(p^\prime,p)$ can have the NG pole:
\begin{eqnarray}
\label{vertex_transveral_nonvanish}
\tilde\Gamma^\mu(p^\prime,p) &=& 2 x g T^\mu(p^\prime,p) \frac{\boldsymbol{\Sigma}_{p^\prime}-\boldsymbol{\Sigma}_{p}}{p^{\prime 2} - p^{2}} \gamma_5
\,,
\hspace{2em}
\end{eqnarray}
where we defined the transversal quantity
\begin{eqnarray}
T^\mu(p^\prime,p) &=& \frac{q^\mu}{q^2}[\slashed{q},\slashed{q}^\prime] - [\gamma^\mu,\slashed{q}^\prime]
\end{eqnarray}
and $q^\prime = p^\prime + p$. Notice the appearance of the parameter $x$, which is now in principle arbitrary real constant. The vertex \eqref{vertex_transveral_nonvanish} does not contain any other poles that the NG pole at $q^2 \rightarrow 0$: The potential pole at $p^{\prime 2} \rightarrow p^{2}$ is canceled by zero in numerator as $\boldsymbol{\Sigma}_{p^\prime} \rightarrow \boldsymbol{\Sigma}_{p}$ and for the fraction one just obtains $\boldsymbol{\Sigma}_{p}^\prime$.


In fact, the new term \eqref{vertex_transveral_nonvanish} should have been considered already in \cite{Pagels:1979hd}, as it is fully consistent with the lowest order of the therein defined \qm{dynamical perturbation theory}. Put another way, the whole Ansatz, including the new term \eqref{vertex_transveral_nonvanish}, is constructed according the same principle, which is linearity in $\boldsymbol{\Sigma}_{p}$ and $\boldsymbol{\Sigma}_{p^\prime}$, linearity in $T_A = g \gamma_5$, no poles other than the NG ones and satisfaction of the WT identity. Mechanical application of these requirement leads straightforwardly to Ansatz \eqref{vertex} with $\tilde\Gamma^\mu(p^\prime,p)$ given by \eqref{vertex_transveral_nonvanish}. The new term is not singled out by WT identity either: After all, the particular separation \eqref{vertex} of the Ansatz into the part which saturates the WT identity and the transversal part is from the point of view of the WT identity arbitrary and therefore unphysical. Thus, the new term \eqref{vertex_transveral_nonvanish} is equally well justified as the old terms in \eqref{vertex} and it is mandatory to include it.

Upon accepting the non-vanishing value \eqref{vertex_transveral_nonvanish} of $\tilde\Gamma^\mu(p^\prime,p)$ the factor $(1-12x)$ appears at the derivative term in the numerator of the PS formula \eqref{pagels_stokar} for $F_\pi^2$. Thus, all we obtained in this simple Abelian model from generalizing the traditional approach by taking \eqref{vertex_transveral_nonvanish} is merely an ambiguity in $F_\pi^2$, with no clue which value of $x$ should be the preferred one. In the following section we will see that considering a more complicated non-Abelian theory will allow us to argue on physical grounds in favor of a particular (non-vanishing) value of $x$.

\section{Electroweak gauge bosons}
\label{sec_ew}

Let us now consider the electroweak theory with $n$ generations of the standard $\group{SU}(2)_{\mathrm{L}} \times \group{U}(1)_{\mathrm{Y}}$ fermion multiplets and $m$ singlets, a.k.a.~the right-handed neutrinos.

We first assume that some dynamics, which we do not need to specify, generates the fermion self-energies, breaking the electroweak symmetry $\group{SU}(2)_{\mathrm{L}} \times \group{U}(1)_{\mathrm{Y}}$ down to the electromagnetic $\group{U}(1)_{\mathrm{em}}$. For the charged fermions $f = f_L + f_R$ we assume the self-energies $-\I\boldsymbol{\Sigma}_f = \langle f \bar f\rangle_{\mathrm{1PI}}$ to have the form \eqref{Sigma_form}: 
$\boldsymbol{\Sigma}_f = \Sigma_f^\dag P_L +  \Sigma_f^{\phantom{\dag}} P_R$, $f=u,d,e$, where $\Sigma_f$ are complex $p^2$-dependent $n \times n$ matrices.\footnote{In this section we will usually suppress the momentum dependence of the self-energies and assume it implicitly.}

For the neutrinos one must take into account the possibility that they can acquire not only the Dirac, but also the Majorana self-energies. This is most conveniently accomplished by putting the neutrinos of both chiralities into the same multiplet, the Nambu--Gorkov doublet $\Psi_\nu = \bigl(\begin{smallmatrix} \nu_L + (\nu_L)^\C \\ \nu_R + (\nu_R)^\C \end{smallmatrix} \bigr)$ 
and by considering the self-energy $-\I\boldsymbol{\Sigma}_{\Psi_\nu} = \langle \Psi_\nu \bar \Psi_\nu \rangle_{\mathrm{1PI}}$ again of the form \eqref{Sigma_form}: $\boldsymbol{\Sigma}_{\Psi_\nu} = \Sigma_{\Psi_\nu}^\dag P_L + \Sigma_{\Psi_\nu}^{\phantom{\dag}} P_R$. Here $\Sigma_{\Psi_\nu}$ is a complex $(n+m) \times (n+m)$ $p^2$-dependent matrix, which is symmetric due to $\Psi_\nu^\C = \Psi_\nu$. The point is that it naturally contains the Dirac self-energy $-\I \Sigma_{\nu D} P_R = \langle \nu_L \bar \nu_R \rangle_{\mathrm{1PI}}$, as well as the two Majorana self-energies $-\I \Sigma_{\nu L} P_R = \langle \nu_L (\bar \nu_L)^\C \rangle_{\mathrm{1PI}}$ and $-\I \Sigma_{\nu R} P_R = \langle (\nu_R)^\C \bar \nu_R \rangle_{\mathrm{1PI}}$: $\Sigma_{\Psi_\nu} = \bigl(\begin{smallmatrix} \Sigma_{\nu L} & \Sigma_{\nu D} \\ \Sigma_{\nu D}^\T & \Sigma_{\nu R} \end{smallmatrix} \bigr)$,
where the matrices $\Sigma_{\nu D}$, $\Sigma_{\nu L}$, $\Sigma_{\nu R}$ have the dimensions $n \times m$, $n \times n$, $m \times m$, respectively, and $\Sigma_{\nu L}$ and $\Sigma_{\nu R}$ are symmetric.

To proceed further, we need the dressed vertices of the fermion--anti-fermion pairs with the electroweak gauge bosons. They can be derived by similar reasoning as in the previous Abelian case, i.e., by demanding satisfaction of the WT identities, correct transformation properties under both the continuous and discrete symmetries, Hermiticity, linearity in the fermion self-energies (evaluated only in $p^\prime$ and $p$), linearity in symmetry generators and gauge coupling constants and no poles but the NG ones with correct residues. Again we end up with vertices which are unique up to the ambiguous transversal parts of the type \eqref{vertex_transveral_nonvanish}, proportional to the yet undetermined constant $x$.

For the sake of illustration, we present here explicitly only the quark vertices; the lepton vertices would be notationally more complicated due to the Majorana character of neutrinos, but otherwise completely analogous. In the physical $(\gamma,Z,W^+,W^-)$ basis the vertices read
\begin{eqnarray}
\Gamma_{\gamma}^\mu(p^\prime,p) &=& e Q_f \bigg\{\gamma^\mu - \frac{q^{\prime\mu}}{q \cdot q^\prime} \boldsymbol{\Sigma}_{f-} \bigg\} \,,
\end{eqnarray}
\begin{widetext}
\begin{eqnarray}
\label{Gamma_Z}
\Gamma_{Z}^\mu(p^\prime,p) &=& \frac{g}{2 \cos \theta_{\mathrm{W}}}
\bigg\{ 
\gamma^\mu (v_f - t_{3f} \gamma_5) + \frac{q^\mu}{q^2} t_{3f} \boldsymbol{\Sigma}_{f+} \gamma_5 - \frac{q^{\prime\mu}}{q \cdot q^\prime} v_f \boldsymbol{\Sigma}_{f-} 
- 2 x \frac{T^\mu(p^\prime,p)}{q \cdot q^\prime} t_{3f} \boldsymbol{\Sigma}_{f-} \gamma_5
\bigg\} \,,
\\
\label{Gamma_Wpm}
\Gamma_{W^+}^\mu(p^\prime,p) &=& \frac{g}{\sqrt{2}} 
\bigg\{
\gamma^\mu P_L - \frac{1}{2} \frac{q^\mu}{q^2} \big(\boldsymbol{\Sigma}_{d+}P_L-\boldsymbol{\Sigma}_{u+}P_R\big) - \frac{1}{2} \frac{q^{\prime\mu}}{q \cdot q^\prime} \big(\boldsymbol{\Sigma}_{d-}P_L+\boldsymbol{\Sigma}_{u-}P_R\big)
+x \frac{T^\mu(p^\prime,p)}{q \cdot q^\prime} \big(\boldsymbol{\Sigma}_{d-}P_L-\boldsymbol{\Sigma}_{u-}P_R\big)
\bigg\} 
\nonumber \\ &&
\end{eqnarray}
\end{widetext}
and $\Gamma_{W^-}^\mu(p^\prime,p) = \gamma_0\Gamma_{W^+}^{\mu\dag}(p,p^\prime)\gamma_0$. We denoted $\boldsymbol{\Sigma}_{f\pm} = \boldsymbol{\Sigma}_{f}(p^{\prime 2}) \pm \boldsymbol{\Sigma}_{f}(p^{2})$ and $v_f = t_{3f} - 2 Q_f \sin^2 \theta_{\mathrm{W}}$.

As a net result we obtain the gauge boson mass matrix in the block diagonal form 
\begin{eqnarray}
M^2 &=& \left(\begin{array}{cc} M^2_{W^\pm} & 0 \\ 0 & M^2_{Z\gamma} \end{array}\right) \,,
\end{eqnarray}
where in the $(A_1^\mu,A_2^\mu,A_3^\mu,B^\mu)$ basis we have
\begin{eqnarray}
\label{mass_matrix_W}
M^2_{W^\pm} &=& \left(\begin{array}{cc} g^2 & 0 \\ 0 & g^2 \end{array}\right) 
\big[ F_\pm^2 + (1+12x) \tilde F_\pm^2 \big]
\nonumber \\ && {}
+\left(\begin{array}{cc} 0 & g^2 \\ -g^2 & 0 \end{array}\right) 
(1+12x) \tilde G_\pm^2 \,,
\\
\label{mass_matrix_Z}
M^2_{Z\gamma}  &=&
\left(\begin{array}{cc} g^2 & -gg^\prime \\ -gg^\prime & g^{\prime 2} \end{array}\right) 
\big[ F_0^2 + (1+12x) \tilde F_0^2 \big] \,.
\end{eqnarray}
The form factors $F_\pm^2$, $\tilde F_\pm^2$, $\tilde G_\pm^2$, $F_0^2$, $\tilde F_0^2$ are independent of $g$, $g^\prime$ and $x$ and are in general non-vanishing. And as can be seen from their explicit forms (some of which will be shown below), they are also real, regardless of the precise form of the fermion self-energies, so that the whole mass matrix is real.

Now we can spot the problem: The mass matrix \eqref{mass_matrix_W} for the $W^\pm$ bosons is not symmetric! This is indeed a pathological situation, since a gauge boson mass matrix must be, in any case, of the manifestly symmetric form $M^2_{ab} = F_{aA}F_{bA}$ (where the sum runs over the indices of the broken generators and coupling constants were included in $F_{aA}$). This non-symmetricity of \eqref{mass_matrix_W} thus suggests internal inconsistency of the vertex Ansatz, used for calculating the polarization tensor.

The non-symmetric part of \eqref{mass_matrix_W} is proportional to form factor $\tilde G_\pm^2$, given explicitly as
\begin{eqnarray}
\label{Gpm}
\tilde G_\pm^2    &=& 
\frac{1}{8} N_c \int\!\frac{\d^4 p}{(2\pi)^4} p^2 
\Tr \bigg\{
D_{dL}^{\phantom{\dag}} \Big( D_{uL}^{\phantom{\dag}} \Sigma_u^{\phantom{\dag}} \Sigma_u^{\dag\prime} -  \Sigma_u^\prime \Sigma_u^{\dag} D_{uL}^{\phantom{\dag}}\Big) 
\nonumber \\ && \hspace{-2em} {}
- (u \leftrightarrow d) \bigg\} 
+ \mbox{(similar contribution of leptons)}
\,,
\end{eqnarray}
where $N_c = 3$ and 
$D_{fL}$ denotes
\begin{subequations}
\label{Df}
\begin{eqnarray}
D_{fL} &=& \big(p^2 - \Sigma_f^{\phantom{\dag}} \Sigma_f^{\dag}\big)^{-1} \,.
\end{eqnarray}
The subscript $L$ is here only to distinguish it from the related, yet different quantity $D_{fR}$:
\begin{eqnarray}
D_{fR} &=& (p^2 - \Sigma_f^{\dag} \Sigma_f)^{-1} \,.
\end{eqnarray}
\end{subequations}

From \eqref{Gpm} we can see first of all that $G_\pm^2$ is indeed real: The matrix under the trace is anti-Hermitian, so that the trace is purely imaginary, while another imaginary unit $\I$ comes from $\d^4 p$ via the Wick rotation, so that the integrand is a real function. More importantly, however, we can see that $G_\pm^2$ indeed, in general, does not vanish. In fact, there are special cases when $G_\pm^2$ does vanish: E.g., when the self-energies $\Sigma_u$, $\Sigma_d$ are not complex matrices, but real numbers. But for generic self-energies $G_\pm^2$ does not vanish.

Now the free constant $x$ comes into play. By setting
\begin{eqnarray}
\label{a4}
x &=& - \frac{1}{12}
\end{eqnarray}
we can get rid of the unwanted antisymmetric part of $M^2_{W^\pm}$, irrespective of the actual value of $\tilde G_\pm^2$. The $W^\pm$ and $Z$ masses then read
\begin{subequations}
\begin{eqnarray}
m^2_{W}    &=&  g^2 F_\pm^2  \,, \\
m^2_Z      &=& (g^2+g^{\prime 2}) F_0^2 \,.
\end{eqnarray}
\end{subequations}
The form factors can be expressed as $F_\pm^2 = \mu^2_{q}+\mu^2_{\ell}$ and $F_0^2 = \mu^2_{u}+\mu^2_{d}+\mu^2_{\nu}+\mu^2_{e}$.
The particular contributions to $F_0^2$ are explicitly
\begin{widetext}
\begin{eqnarray}
\label{mu_ud}
\mu^2_{f} &=& -\I\frac{1}{2} N_c \int\!\frac{\d^4 p}{(2\pi)^4}\,\Tr
\bigg\{
\Big[
\big(\Sigma_f^{\vphantom{\dag}}\Sigma_f^\dag\big) - \frac{1}{2} p^2
\big(\Sigma_f^{\vphantom{\dag}}\Sigma_f^\dag\big)^\prime
\Big]
D_{fL}^{2\vphantom{\dag}}
\bigg\} \,,
\\
\label{mu_nu}
\mu^2_{\nu} &=&
-\I\frac{1}{2} \int\!\frac{\d^4 p}{(2\pi)^4}\,\Tr\bigg\{
\Big[
\big(\Sigma_{\nu D}^{\vphantom{\dag}}\Sigma_{\nu D}^\dag\big) - \frac{1}{2}p^2
\big(\Sigma_{\nu D}^{\vphantom{\dag}}\Sigma_{\nu D}^\dag\big)^\prime
\Big]D_{\nu L}^2
\nonumber \\ && \hspace{6.1em} {}
+2\Big[
\big(\Sigma_{\nu L}^{\vphantom{\dag}}\Sigma_{\nu L}^\dag\big) - \frac{1}{2}p^2
\big(\Sigma_{\nu L}^{\vphantom{\dag}}\Sigma_{\nu L}^\dag\big)^\prime
\Big]D_{\nu L}^2
\nonumber \\ && \hspace{6.1em} {}
+\frac{1}{2}
\bigg(3
\Big[
\big(\Sigma_{\nu D}^{\vphantom{\dag}\T}\Sigma_{\nu L}^{\dag}\big) - \frac{1}{2}p^2
\big(\Sigma_{\nu D}^{\vphantom{\dag}\T}\Sigma_{\nu L}^{\dag}\big)^\prime
\Big]D_{\nu L}^{\vphantom{\dag}}D_{\nu M}^{\vphantom{\dag}}
-\frac{1}{2}p^2
\big(\Sigma_{\nu D}^{\vphantom{\dag}\T}\Sigma_{\nu L}^{\dag}\big)
\big(D_{\nu L}^{\vphantom{\prime}}D_{\nu M}^{\prime}-D_{\nu L}^{\prime}D_{\nu M}^{\vphantom{\prime}}\big)
+ \hc
\bigg)
\nonumber \\ && \hspace{6.1em} {}
+\frac{1}{2}
\bigg(
\Big[
\big(\Sigma_{\nu R}^{\vphantom{\dag}}\Sigma_{\nu D}^\dag\big) - \frac{1}{2}p^2
\big(\Sigma_{\nu R}^{\vphantom{\dag}}\Sigma_{\nu D}^\dag\big)^\prime
\Big]D_{\nu L}^{\vphantom{\dag}}D_{\nu M}^{\vphantom{\dag}}
-\frac{1}{2}p^2
\big(\Sigma_{\nu R}^{\vphantom{\dag}}\Sigma_{\nu D}^\dag\big)
\big(D_{\nu L}^{\vphantom{\prime}}D_{\nu M}^{\prime}-D_{\nu L}^{\prime}D_{\nu M}^{\vphantom{\prime}}\big)
+ \hc
\bigg)
\bigg\} \,,
\nonumber \\ &&
\end{eqnarray}
where $N_c = 3$ for $f=u,d$ and $N_c = 1$ for $f=e$, and the contributions to $F_\pm^2$ are
\begin{eqnarray}
\label{mu_q}
\mu^2_{q} &=&
-\I\frac{1}{2} N_c \int\!\frac{\d^4 p}{(2\pi)^4}\, \Tr\bigg\{
\Big[
\big(\Sigma_u^{\vphantom{\dag}}\Sigma_u^\dag\big) - \frac{1}{2} p^2
\big(\Sigma_u^{\vphantom{\dag}}\Sigma_u^\dag\big)^\prime
\Big]
D_{dL}D_{uL}
-\frac{1}{2} p^2 \big(\Sigma_u^{\vphantom{\dag}}\Sigma_u^\dag\big)
\Big[
D_{dL}^{\vphantom{\prime}}D_{uL}^{          \prime} -
D_{dL}^{          \prime} D_{uL}^{\vphantom{\prime}}
\Big]
\nonumber \\ && \hspace{7.8em} {}
+
\Big[
\big(\Sigma_d^{\vphantom{\dag}}\Sigma_d^\dag\big) - \frac{1}{2} p^2
\big(\Sigma_d^{\vphantom{\dag}}\Sigma_d^\dag\big)^\prime
\Big]
D_{uL}D_{dL}
-\frac{1}{2} p^2 \big(\Sigma_d^{\vphantom{\dag}}\Sigma_d^\dag\big)
\Big[
D_{uL}^{\vphantom{\prime}}D_{dL}^{          \prime} -
D_{uL}^{          \prime} D_{dL}^{\vphantom{\prime}}
\Big]
\bigg\} \,,
\\
\label{mu_ell}
\mu^2_{\ell} &=&
-\I\frac{1}{2} \int\!\frac{\d^4 p}{(2\pi)^4}\,\Tr\bigg\{
\Big[
\big(\Sigma_{\nu D}^{\vphantom{\dag}}\Sigma_{\nu D}^\dag\big) - \frac{1}{2}p^2
\big(\Sigma_{\nu D}^{\vphantom{\dag}}\Sigma_{\nu D}^\dag\big)^\prime
\Big]
D_{eL}^{\vphantom{\dag}}D_{\nu L}^{\vphantom{\dag}}
- \frac{1}{2}p^2
\big(\Sigma_{\nu D}^{\vphantom{\dag}}\Sigma_{\nu D}^\dag\big)
\Big[
D_{eL}^{\vphantom{\prime}}D_{\nu L}^{\prime} - D_{eL}^{\prime}D_{\nu L}^{\vphantom{\prime}}
\Big]
\nonumber \\ && \hspace{6.6em} {} 
+
\Big[
\big(\Sigma_{\nu L}^{\vphantom{\dag}}\Sigma_{\nu L}^\dag\big) - \frac{1}{2}p^2
\big(\Sigma_{\nu L}^{\vphantom{\dag}}\Sigma_{\nu L}^\dag\big)^\prime
\Big]
D_{eL}^{\vphantom{\dag}}D_{\nu L}^{\vphantom{\dag}}
- \frac{1}{2}p^2
\big(\Sigma_{\nu L}^{\vphantom{\dag}}\Sigma_{\nu L}^\dag\big)
\Big[
D_{eL}^{\vphantom{\prime}}D_{\nu L}^{\prime} - D_{eL}^{\prime}D_{\nu L}^{\vphantom{\prime}}
\Big]
\nonumber \\ && \hspace{6.6em} {} 
+
\Big[
\big( \Sigma_{\nu D}^{\T\vphantom{\dag}}\Sigma_{\nu L}^\dag \big) - \frac{1}{2}p^2
\big( \Sigma_{\nu D}^{\T\vphantom{\dag}}\Sigma_{\nu L}^\dag \big)^\prime
\Big]
D_{eL}^{\vphantom{\dag}}D_{\nu M}^{\vphantom{\dag}}
- \frac{1}{2}p^2
\big( \Sigma_{\nu D}^{\T\vphantom{\dag}}\Sigma_{\nu L}^\dag \big)
\Big[
D_{eL}^{\vphantom{\prime}}D_{\nu M}^{\prime} - D_{eL}^{\prime}D_{\nu M}^{\vphantom{\prime}}
\Big]
\nonumber \\ && \hspace{6.6em} {} 
+
\Big[
\big( \Sigma_{\nu R}^{\vphantom{\dag}}\Sigma_{\nu D}^\dag \big) - \frac{1}{2}p^2
\big( \Sigma_{\nu R}^{\vphantom{\dag}}\Sigma_{\nu D}^\dag \big)^\prime
\Big]
D_{eL}^{\vphantom{\dag}}D_{\nu M}^{\vphantom{\dag}}
- \frac{1}{2}p^2
\big( \Sigma_{\nu R}^{\vphantom{\dag}}\Sigma_{\nu D}^\dag \big)
\Big[
D_{eL}^{\vphantom{\prime}}D_{\nu M}^{\prime} - D_{eL}^{\prime}D_{\nu M}^{\vphantom{\prime}}
\Big]
\nonumber \\ && \hspace{6.6em} {} 
+
\Big[
\big( \Sigma_{e}^{\vphantom{\dag}}\Sigma_{e}^\dag \big) - \frac{1}{2}p^2
\big( \Sigma_{e}^{\vphantom{\dag}}\Sigma_{e}^\dag \big)^\prime
\Big]
D_{\nu L}^{\vphantom{\dag}}D_{eL}^{\vphantom{\dag}}
- \frac{1}{2}p^2
\big( \Sigma_{e}^{\vphantom{\dag}}\Sigma_{e}^\dag \big)
\Big[
D_{\nu L}^{\vphantom{\prime}}D_{eL}^{\prime} - D_{\nu L}^{\prime}D_{eL}^{\vphantom{\prime}}
\Big]
\bigg\} \,,
\end{eqnarray}
\end{widetext}
where the prime is derivative with respect to $p^2$.
We used here, besides the notation $\eqref{Df}$ for propagators of the charged fermions,
a similar notation for the propagator of neutrinos: 
\begin{eqnarray}
\left(\begin{array}{cc} D_{\nu L} & D_{\nu M} \\ D_{\nu M}^\dag & D_{\nu R}^\T \end{array}\right) &=& (p^2-\Sigma_{\Psi_\nu}^{\phantom{\dag}}\Sigma_{\Psi_\nu}^\dag)^{-1}
\,.
\end{eqnarray}
The dimensions of the blocks $D_{\nu L}$, $D_{\nu R}$ and $D_{\nu M}$ correspond to those of blocks of $\Sigma_{\Psi_\nu}$.

Notice that the form factors $F_\pm^2$ and $F_0^2$ are indeed real: All traces are over Hermitian matrices and therefore are real, while the imaginary units $\I$ in front of the integrals are canceled by another $\I$ from the Wick rotation.


\section{Comparison with literature}
\label{sec_comparison}

While the formulae \eqref{mu_nu}, \eqref{mu_ell} for contributions $\mu^2_\nu$, $\mu^2_\ell$ including the Majorana neutrinos are genuine new, the formulae analogous to \eqref{mu_ud}, \eqref{mu_q} for quarks contributions $\mu^2_{f=u,d}$, $\mu^2_q$ have already been derived in the literature \cite{Miransky:1988xi}. They differ from our results by the coefficients at the derivative terms $(\Sigma_f^{\phantom{\dag}}\Sigma_f^\dag)^\prime$, which are twice as small than in \eqref{mu_ud}, \eqref{mu_q}. This discrepancy is to be attributed to the value of the parameter $x$: While we set the non-vanishing value \eqref{a4}, the authors of \cite{Miransky:1988xi} effectively set $x = 0$ by ignoring the terms of the type \eqref{vertex_transveral_nonvanish} in the vertices. The reason why they did not found any inconsistency regarding the non-symmetricity of the gauge boson mass matrix (or different masses of the $W^+$ and $W^-$ bosons) was that they considered an oversimplified case of only one fermion generation and real quark self-energies $\Sigma_{u}$, $\Sigma_{d}$. Had they considered a more general case of $\Sigma_{u}$, $\Sigma_{d}$ being either \emph{complex} or \emph{matrices} (or both), they would have obtained a non-symmetric mass matrix for $W^\pm$ bosons, as can be seen from \eqref{Gpm}.

In \cite{Blumhofer:1993kv} the masses of $W^\pm$ and $Z$ boson are obtained by calculating not the whole transversal polarization tensor, but only its $g^{\mu\nu}$ part. Besides assuming also only one fermion generation and real quark self-energies, the most crude assumption is setting $g^\prime = 0$. Interestingly enough, the authors of \cite{Blumhofer:1993kv} still do obtain the formulae equivalent (via integration by parts) to ours for $\mu^2_{f=u,d}$, $\mu^2_q$, including the correct factors of $1/2$ at the derivative terms. However, their method cannot be applied to more general cases like, e.g., theories with different gauge groups, and also upon relaxing the assumption $g^\prime = 0$ it yields wrong results like, e.g., a massive photon.

It is natural to assume that the non-vanishing value \eqref{a4} of the parameter $x$, derived in the context of electroweak interaction, should apply also to the Abelian model from Sec.~\ref{sec_review}. Using this value we obtain the corrected PS formula
\begin{eqnarray}
\label{pagels_stokar_correct}
F_\pi^2 &=& -2 \I \int\!\frac{\d^4 p}{(2\pi)^4} 
\frac{|\Sigma_p|^2 - \frac{1}{2}p^2|\Sigma_p|^{2\prime}}{(p^2-|\Sigma_p|^2)^2}
\,.
\end{eqnarray}
Notice that the difference from the original PS formula \eqref{pagels_stokar} is again just the factor of $1/2$ at the derivative term.

Pagels and Stokar used in \cite{Pagels:1979hd} their formula \eqref{pagels_stokar} to estimate the value of $F_\pi$ in QCD. They adopted
the Ansatz $\Sigma_p = 4 m^3/p^2$ with the constituent quark mass $m=244 \MeV$.
From their formula \eqref{pagels_stokar} they obtained (for two quark flavors) the estimate $F_\pi = 83 \MeV$, which was surprisingly close to the experimental value $F_\pi = 93 \MeV$.
Had they used rather the corrected formula \eqref{pagels_stokar_correct}, they would have obtained $F_\pi = 96 \MeV$, i.e., the agreement would have been actually even better.

\section{Generalization}
\label{sec_generalization}

The analysis of this letter can be generalized to arbitrary gauge group $\group{G}$, spontaneously broken by fermion self-energies down to some subgroup $\group{H} \subseteq \group{G}$. The technical details will be presented elsewhere, but the conclusions are the same: The resulting gauge boson mass matrix is again generally non-symmetric. And again, this can be fixed once one takes into account the (appropriate generalization of the) transversal term \eqref{vertex_transveral_nonvanish}, with the constant $x$ set again to the \emph{same} value \eqref{a4}.

This suggests that results of Sec.~\ref{sec_ew}, i.e., the existence and necessity of the terms of the type \eqref{vertex_transveral_nonvanish} with the particular value \eqref{a4} of $x$, are not special to the electroweak theory, but are more general and common to much larger class of spontaneously broken gauge theories.

\section{Conclusions}
\label{sec_conclusions}

In this letter we made two basic observations, previously unnoticed in the literature. First, we showed that the Ansatz for the proper vertex function $\langle A^\mu \psi \bar \psi \rangle$ (used in the PS treatment of spontaneously broken gauge symmetry) can be augmented by the new term of the type \eqref{vertex_transveral_nonvanish}. Second, we found that using the traditional PS treatment (without the new term in the vertex Ansatz) of the electroweak symmetry breaking one arrives at non-symmetric gauge boson mass matrix (unless one considers unrealistic case of only one fermion generation with real self-energies).

We demonstrated that these two observations can be naturally combined: Upon considering the new term \eqref{vertex_transveral_nonvanish} in vertex Ansatz (with unique value of the otherwise free parameter $x$) the electroweak gauge boson mass matrix is made manifestly symmetric, without assuming any special form of the fermion self-energies. In other words, the new term \eqref{vertex_transveral_nonvanish} is not only allowed, it is in fact necessary to make the theory (or the vertex Ansatz) consistent.

As a result we provided explicit formulae for the $W^\pm$ and $Z$ masses in terms of the fermion self-energies. In comparison with the similar formulae in the literature we have achieved two improvements. First, due to the new term in the vertex Ansatz, we managed to generalize them to arbitrary number of fermion generations. Second, we considered, for the first time, also the most general neutrino setup with arbitrary number of right-handed neutrinos, together with left- and right-handed Majorana self-energies.  For both the electrically charged fermions and the neutrinos we also allowed for arbitrary mixing.

%
%

Finally, we have also found a correction to the PS formula for the pion decay constant in QCD, again by assuming the non-vanishing value \eqref{a4} of $x$.


\begin{acknowledgments}
I would like to thank Ji\v{r}\'{\i} Ho\v{s}ek and Adam Smetana for useful comments. The work was supported by TJ~Balvan Praha.
\end{acknowledgments}




%
%
%
%
%
%
%
%
%

%

\end{document}